# Quantum critical scaling at the edge of Fermi liquid stability in a cuprate superconductor

N. P. Butch*, K. Jin*, K. Kirshenbaum, R. L. Greene, and J. Paglione

*Center for Nanophysics & Advanced Materials and Department of Physics,*
*University of Maryland, College Park, MD 20742, USA*

In the high temperature cuprate superconductors, the pervasiveness of anomalous electronic transport properties suggests that violation of conventional Fermi liquid behavior is closely tied to superconductivity[1,2]. In other classes of unconventional superconductors, atypical transport is well correlated with proximity to a quantum critical point[3,4], but the relative importance of quantum criticality in the cuprates[5,6] remains uncertain. Here we identify quantum critical scaling in the electron-doped cuprate material $La_{2-x}Ce_xCuO_4$ (LCCO) with a line of quantum critical points that surrounds the superconducting phase as a function of magnetic field and charge doping. This zero-temperature phase boundary, which delineates a metallic Fermi liquid regime from an extended non-Fermi liquid ground state, closely follows the upper critical field of the overdoped superconducting phase and gives rise to an expanse of distinct non Fermi liquid behavior at finite temperatures. Together with signatures of two distinct flavors of quantum fluctuations, this suggests that quantum criticality plays a significant role in shaping the anomalous properties of the cuprate phase diagram.

A longstanding issue in the quest to understand high-temperature superconductivity in the cuprates is in regard to the nature of the underlying ground state[7]. The recent observations of quantum oscillations in underdoped $YBa_2Cu_3O_{6+x}$ (ref. 8) have provided a significant advance to our knowledge of the progression of this ground state through the cuprate phase diagram. The presence of small Fermi surface (FS) pockets distinct from the large FS structure observed in overdoped cuprates[9] requires the existence of a FS reconstruction, which logically occurs at a quantum phase transition between ground states that modify the symmetry of the Brillouin zone. With the origin of superconductivity still under hot debate, how the phase diagram is "shaped" by the evolution of these ground states remains as a crucial question.

A FS transformation has also been directly observed in the electron-doped cuprates as a function of doping, for instance as observed in $Nd_{2-x}Ce_xCuO_4$ (refs 10, 11). This evolution is consistent with several indications of a quantum critical point associated with the suppression of antiferromagnetic order near optimal doping for superconductivity, and the appearance of a Fermi liquid (FL) ground state on the overdoped side. With relatively low upper critical field values, the electron-doped cuprates allow for a unique opportunity to study the underlying ground state of the phase diagram in much detail[12]. LCCO is particularly unique in that its superconducting (SC) "dome" is centered at relatively lower Ce concentrations[13], making it possible to study the complete suppression of superconductivity by both doping and magnetic field. More important, in LCCO the selective response of spin fluctuations and superconductivity to magnetic field and doping allow us to segregate what appears to be a complicated mixture of behaviors into two distinct signatures of criticality.

As shown in Fig.1, the non-superconducting FL ground state of overdoped LCCO can be readily accessed by either of two ways: doping in electrons beyond a critical value $x_c$, or increasing magnetic field above a critical value $B_c$. Both tuning parameters suppress superconductivity and induce a FL ground state that appears to emerge continuously beyond a series of quantum critical points that form a continuous line along the ground state ($T = 0$) plane. A direct signature of this criticality, i.e., critical divergence as a function of an experimental tuning parameter[14], is found as a function of magnetic field $B$: upon approach to the critical field $B_c$ from above, a divergence in the quasiparticle-quasiparticle scattering cross-section occurs as the temperature range of Fermi liquid behavior, denoted by $T_{FL}$, is driven to zero at $B_c$. At each doping the quadratic temperature coefficient $A_2$ determined from fits of the form $\Delta\rho = \rho - \rho_0 = A_2 T^2$ in the FL state (Fig. 2), strongly increases with decreasing field magnitude and diverges as a function of field $\Delta B = B - B_c(x)$. Furthermore, the reduced field scale $\Delta B / B_c(x)$ diverges with a universal critical exponent $\alpha = 0.38$ that is the same for all dopings considered (Fig. 3a), indicating that $B_c(x)$ acts as a line of quantum critical points (see Supplemental Information).

---

*These authors contributed equally

Strikingly similar divergences have been identified in several different systems exhibiting magnetic field-tuned quantum criticality, including the heavy-fermion materials $CeCoIn_5$ (ref. 15), $CeAuSb_2$ (ref. 16), $YbRh_2Si_2$ (ref. 17) and $YbAlB_4$ (ref. 18), with critical exponents 1.37, 1.0, 1.0 and 0.50, respectively. In contrast to classical transitions, the sensitivity to effective dimensionality involved in a quantum phase transition can lead to non-universal critical exponents[14]. In LCCO, the observation of a universal exponent at several doping levels is unprecedented, but is limited to magnetic field tuning. When considering doping as a tuning parameter, the system can also be tuned to approach the critical field but with a distinct critical exponent. That is, $A_2$ also scales as a function of reduced doping $\Delta x / x_c(B)$ for different constant magnetic field values, with a critical exponent $\beta = 0.72$ (Fig. 3b). This is a rare example of a material where both magnetic field and doping can drive the electronic system to quantum criticality in a similar but distinct manner. These two tuning parameters, one adding charge carriers and one breaking time reversal symmetry, likely alter the excitation spectrum in fundamentally different ways, as considered in the case of heavy-fermion systems with similar orthogonal tuning parameters[19]. However, they also smoothly connect the ground state boundaries that define the phase diagram on the overdoped side.

In LCCO, resistivity data can be scaled as a function of $\Delta B/T$ as shown in Fig. 3, providing a second key signature of the "reach" of a quantum phase transition. First observed in heavy fermion materials[20], this type of energy-temperature scaling not only indicates a quantum critical system below its upper critical dimension, but also reflects the lack of an energy scale other than temperature itself[14]. In such a case, the transport can be described generally as a function $f(\Delta B^\gamma / T)$ of both field and temperature, with asymptotic limits in both FL ($\Delta\rho \propto T^2$) and NFL ($\Delta\rho \propto T^n$) regions (see Supplemental Information). Through this approach, the anomalous $T^n$ scattering and the magnetic field-tuned divergence of $A_2$ with exponent $\alpha$ are shown to be two aspects of the same critical behavior, with a self-consistency given by $\alpha = \gamma(2 - n)$. A scaling exponent $\gamma$ is obtained for both $x = 0.15$ and 0.17, but with different values of $0.4 \pm 0.1$ for $x = 0.15$ (Fig. 3c) and $1.0 \pm 0.02$ for $x = 0.17$ (Fig. 3d). Given the same measured critical divergence exponent $\alpha = 0.38$ for both dopings, self-consistency requires that the power law exponent $n$ must be different for these two dopings. Upon inspection of the phase diagram of Figure 1, one can see this correspondence is indeed verified: at finite temperatures immediately above the QCP at $B_c(x)$ for each doping, $\Delta\rho \propto T^n$ is best fit with $n=1.0$ for $x = 0.15$, and $n = 1.6$ for $x = 0.17$ (see Supplemental Information Fig. 1), confirming self-consistency.

But what is the origin of these inherently different scattering rate behaviors, with $n=1.0$ and $n=1.6$? One of the most extraordinary characteristics of the cuprates is the hallmark temperature-linear resistivity, which was shown in LCCO to persist over three decades in temperature above $B_c$ ($x = 0.15$), and to have a strong correlation with the pairing strength itself[21]. In LCCO, strong circumstantial evidence indicates that this anomalous scattering rate arises due to an antiferromagnetic quantum critical point that lies deep within the SC dome near $x_{FS} = 0.14$ (refs 22, 23), where the Fermi surface reconstructs as in other electron-doped cuprates[24,25]. Fluctuations emanating from this critical point are likely to be responsible for the $n = 1.0$ power law[26], spawning an extended spin fluctuation (SF) region defined by the $n = 1.0$ scattering behavior that dominates a substantial range of temperature, magnetic field, and doping. Of course, the inception of superconductivity likely consumes much of the entropy associated with such a state[27], filling in most of the SF phase space as shown in Fig.1. However, as shown in Fig.1b, a tantalizing glimpse of a possible non Fermi liquid phase (NFL) may be present between the SC upper critical field $B_{c2}$ and $B_c$, where an extended range of $T = 0$ NFL behavior endures much like in other anomalous systems[28-30].

Thus, at $x = 0.15$, the $n = 1.0$ scattering mechanism is dominant, extending to the zero-temperature limit once $B_{c2}$ is surpassed, and the resultant $\Delta B/T$ scaling obeys the expected self-consistency in a wide range of fields and temperatures reaching up to the SF scale $T_1$. However, upon increasing doping from $x = 0.15$ the SF energy scale is dramatically reduced both in temperature and in field, with both scales terminating at the critical doping $x_c = 0.175$ where both $T_1$ and $B_1$ approach zero together with $T_c$ and $T_{FL}$. Given the intimate correlation between $T_1$ and $T_c$ in zero field[21], the discrepancy between their magnetic field dependence is all the more remarkable. It indicates that magnetic field does not destabilize superconductivity by destroying the mediating spin fluctuations, but rather through more mundane orbital effects. For instance, at $x = 0.15$, the upper temperature limit of the SF region, denoted as $T_1$, is much more robust against magnetic field than $T_c$ itself, extrapolating to a zero-temperature field scale $B_1$ that far surpasses $B_{c2}$ (see Fig.1b). But at higher doping, $T_1$ and $T_c$ are both suppressed at an almost equal rate toward zero close to $B_c$, and the $n = 1.6$ power law characterizes the dominant scattering rate at temperatures directly above the quantum critical point. For instance, in the special case of $x = 0.17$ at 4 T (Fig.1c), this power law persists from the lowest measured temperatures up to 13 K (Figs 1 and 2), showing that $T^{1.6}$ resistivity persists over almost three decades in temperature when it is the dominant scattering mechanism.

This correspondence underscores two major points. First, the magnetic field-induced divergence, critical scaling and the NFL scattering temperature dependence can be understood within a self-consistent framework. Second, the fact that this self-consistency "adjusts" according to which scattering is dominant is evidence for critical behavior arising from two origins – two sets of anomalous scattering, two forms of scaling and self-consistent critical exponents. Clearly, there are two distinct scattering behaviors that respond differently to doping and magnetic field, and the competition of these two scattering mechanisms is directly borne out in the temperature dependence of resistivity throughout the field-doping phase diagram. With the $n = 1.0$ power law likely arising from scattering with fluctuations associated with the antiferromagnetism of the parent compound, the $n=1.6$ power law appears to be a distinct signature of a second type of quantum critical fluctuation. Interestingly, this power law is strikingly similar to that observed in the hole-doped cuprates $La_{2-x}Sr_xCuO_4$ (ref. 1) and $Tl_2Ba_2CuO_{6+x}$ (ref. 31) in the vicinity of $x_c$, suggesting the quantum critical endpoint of the SC phase may give rise to fluctuations that cause this particular anomalous scattering behavior. In fact, recent measurements of both $La_{2-x}Sr_xCuO_4$ (ref. 32) and $Tl_2Ba_2CuO_{6+x}$ (ref. 33) indeed show quantum critical behavior originating from the end of the SC dome, pointing to a universal nature of the quantum phase transition separating the superconducting and Fermi liquid ground states. The possibility of calculating a nonperturbative critical theory of such fluctuations for a disorder-driven SC quantum critical point[34] shows promise for confirming such a scenario.

Clearly, quantum criticality plays a significant role in shaping the phase diagram of the electron-doped cuprates, both in optimizing the superconductivity as well as limiting its extent. The ensuing picture is that two proximal quantum critical points compete in the cuprate phase diagram. The first, positioned near optimal doping, gives rise to spin fluctuations that stabilize unconventional superconductivity. The second, at $B_c(x)$, owes its very existence to the first as it is born of the suppression of superconductivity and the emergence of the normal FL state. The result is a complex but tractable interplay of competing quantum critical fluctuations that conspire to shape the phase diagram that has become the ubiquitous signature of high-temperature superconductivity.

**Methods**

*Samples*: The c-axis-oriented LCCO films were deposited on (100) $SrTiO_3$ substrates by pulsed laser deposition utilizing a KrF excimer laser. The annealing process for each Ce concentration was optimized such that samples showed the narrowest SC transition widths or metallic behavior down to the lowest measured temperature (20 mK), whereas non-optimized samples usually showed an upturn at low temperature, as previously reported. The films were patterned into Hall bar bridges using photolithography and ion milling techniques. Several samples of each concentration were studied to ensure that the data are representative.

*Measurements*: Electrical transport measurements at temperatures greater than 2 K were carried out in a commercial cryostat equipped with a 14 T magnet, while lower temperature measurements down to 20 mK were performed in a dilution refrigerator equipped with a 15 T magnet. Data from the two platforms were measured with overlapping temperature ranges. Current was applied in the *ab*-plane while the magnetic field was applied along the *c*-axis for all the measurements.

**Acknowledgements**

We would like to thank A. Chubukov, V. Galitski, J. Schmalian, L. Taillefer and C.M. Varma for useful discussions. This research was supported by the NSF under DMR-0952716 (J.P. and K.K.) and DMR-0653535 (R.L.G.) and the Maryland Center for Nanophysics and Advanced Materials (K.J. and N.P.B.).

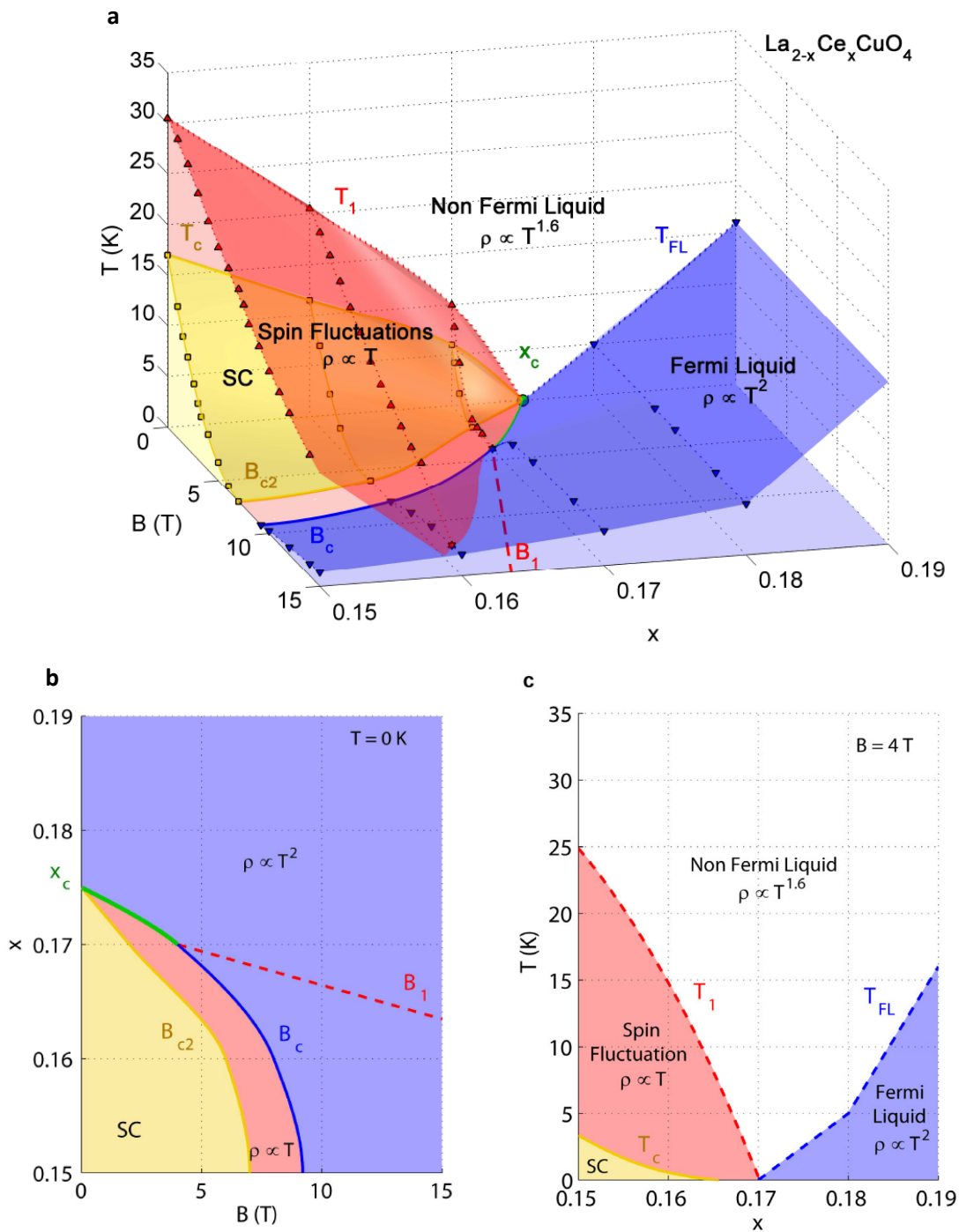

**Figure 1 | Shaping of the overdoped cuprate phase diagram. a**, The interplay between superconducting (SC), spin fluctuation, and Fermi Liquid (FL) phases in La$_{2-x}$Ce$_x$CuO$_4$ near the quantum critical endpoint $x_c$ evolves as a function of electron doping ($x$), magnetic field ($B$), and temperature ($T$). Distinct phase boundaries between SC (yellow) and FL (blue) ground states are determined by a competition of two distinct yet related types of quantum fluctuations that give rise to separable non Fermi liquid behavior,

characterized by $\Delta\rho \propto T$ (red) and $\Delta\rho \propto T^{1.6}$ (white) resistivity temperature dependences. This is found throughout the phase diagram at temperatures above the line of quantum critical points $B_c(x)$ that extends to the zero-field critical doping $x_c$ where the SC critical temperature $T_c$ and crossover temperatures $T_1$ and $T_{FL}$ meet. Unconventional $\sim T^{1.6}$ scattering persists in applied magnetic fields above both the FL and SF regions, but is dominated by a linear-$T$ scattering mechanism in the regime below $T_1$, where SF scattering is dominant. The origin of the SF regime is a quantum critical point at $x = 0.14$ (ref. 21). **b**, The ground state evolution of these phases in the $T = 0$ doping-field plane exhibits a distinct separation between FL and SF ground states, with an extended non Fermi liquid phase (red) characterized by linear-$T$ scattering in the $T=0$ limit. Closer to $x_c$, $T^{1.6}$ behavior dominates and extends to the $T = 0$ limit in a confined region (green line). Although the extrapolated limit of the SF phase $B_1$ (red dashed line) extends to high field, the SC upper critical field $B_{c2}$ and the FL phase boundary $B_c$ restrict the range of the actual SF ground state. Critical scaling behavior is associated with $B_c$ (see text), establishing it as a line of quantum critical points that terminates at $x_c$. **c**, A constant-field cut of the phase diagram at 4 T highlights the region where the SF ground state separates the SC phase from the FL phase and $T^{1.6}$ resistivity extends to zero temperature.

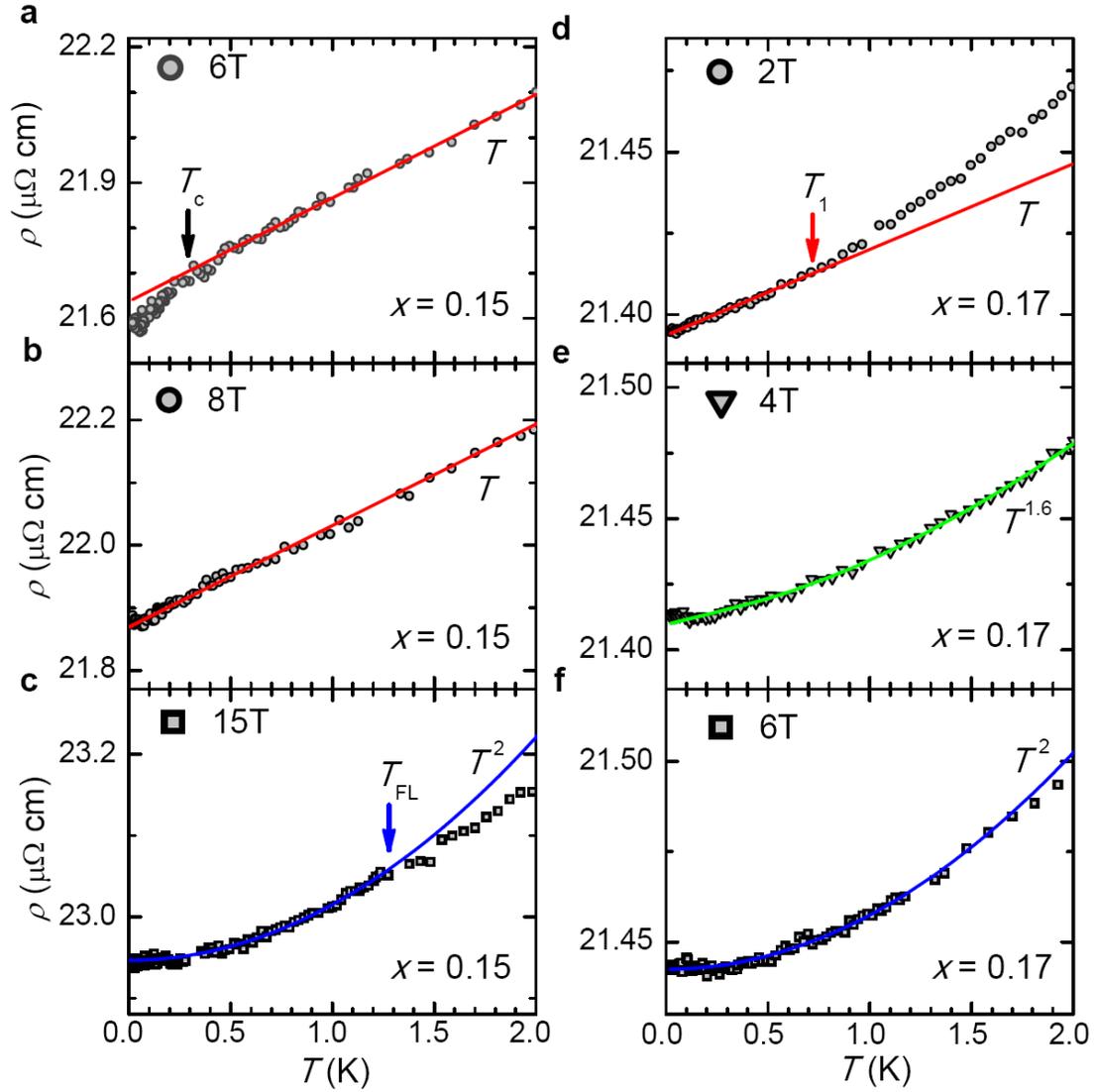

**Figure 2 | Contrasting non-Fermi liquid transport behaviors.** The evolution of the zero-temperature limiting behavior of electrical resistivity $\rho(T)$ for two characteristic superconducting films of $La_{2-x}Ce_xCuO_4$ with $x$ = 0.15 (**a**-**c**) and 0.17 (**d**-**f**) with applied magnetic field demonstrates the isolation of two distinct non Fermi liquid power laws. **a** and **b**, For $x$= 0.15, the suppression of the superconducting state just above 6 T reveals the extension down to the $T$ = 0 limit of the ubiquitous temperature-linear resistivity associated with spin fluctuation scattering[21]. **c**, This behavior is eventually displaced by a Fermi liquid ground state with conventional ~$T^2$ scattering behavior persisting up to a characteristic temperature $T_{FL}$ (blue arrows) at higher fields. In contrast, increasing doping closer to the critical endpoint of the superconducting phase at $x_c$ = 0.175 reveals a different anomalous scattering behavior. **d**, For $x$ = 0.17, the temperature-linear scattering that is present above $T_c$ in a finite range of temperatures up to $T_1$ (red arrows) in zero field is displaced by a more dominant scattering mechanism upon increase of field. **E**, At 4 T, a ~$T^{1.6}$ power law (green line fit) is observed to extend down to zero temperature, and is likely due to fluctuations associated with endpoint of the superconducting phase (see text).

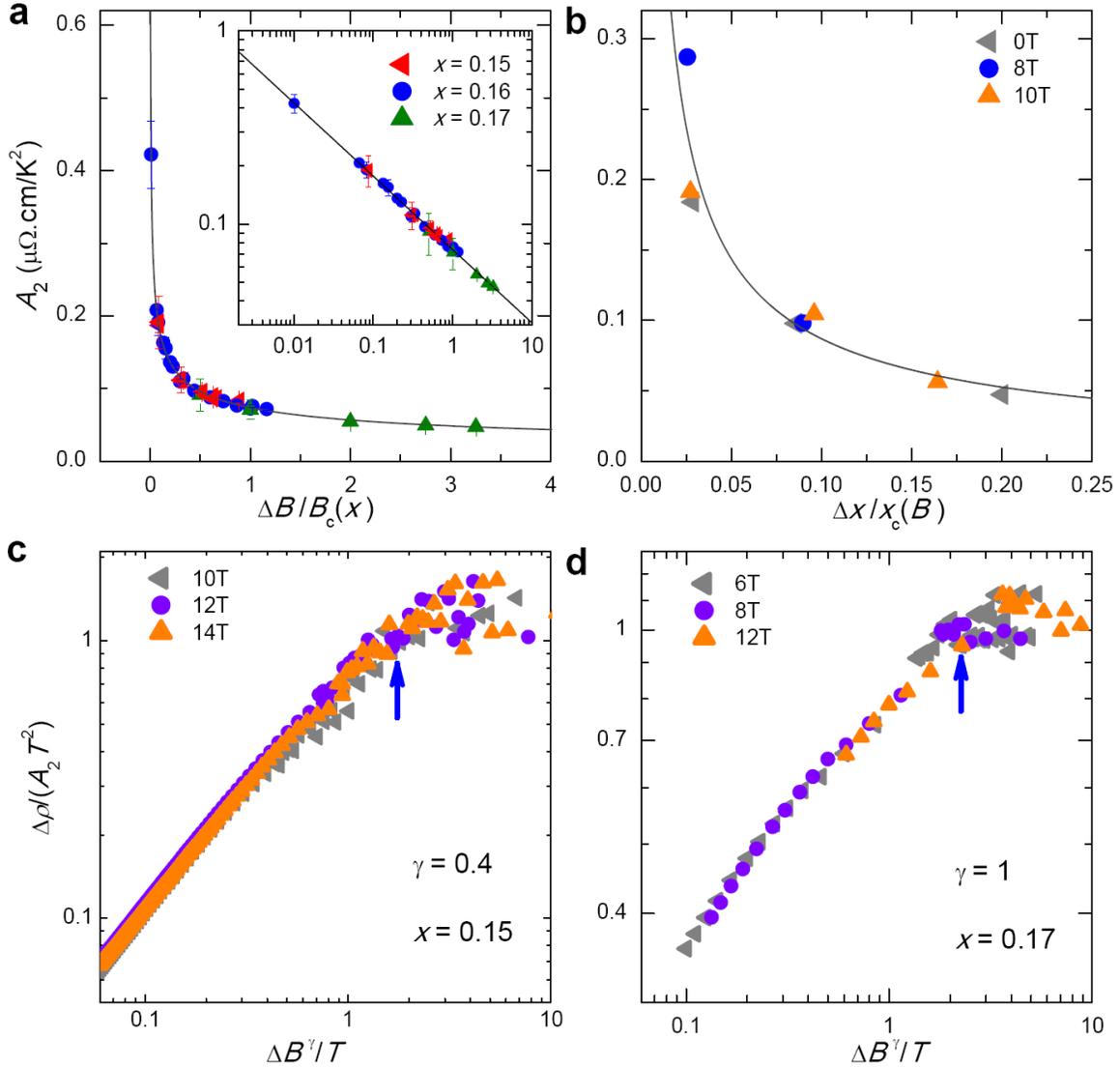

**Figure 3 | Signatures of magnetic field- and charge doping-tuned quantum criticality.** **a,** A strong increase of the quasiparticle-quasiparticle scattering coefficient $A_2$ (from fits of $\rho = \rho_0 + A_2 T^2$) as a function of magnetic field provides evidence for a field-tuned quantum critical point, with a critical divergence observed to occur at the quantum critical field $B_c$ as a function of the field tuning parameter $\Delta B = B - B_c$. Taken in the zero-temperature limit for three Ce concentrations spanning the overdoped region of superconducting LCCO, all of the data fit to one divergent function, $A_2 = A_0(\Delta B/B_c)^{-\alpha}$ with critical exponent $\alpha = 0.38$, indicating that the doping-dependent critical field $B_c(x)$ constitutes a quantum phase transition. The normalizing field-independent factor $A_0$ is equal to unity for $x=0.15$, and scaled to unity for other samples to remove variations due to geometric factor uncertainties, and the inset presents the same data on a log-log plot with slope representative of the same exponent $\alpha$. **b,** A critical divergence in $A_2$ is also witnessed to occur as a function of Ce concentration tuning parameter $\Delta x = x - x_c$ upon approach to the critical doping $x_c$ where the superconducting, Fermi liquid, and spin fluctuation

phases terminate. Data for different magnetic field values are fitted by $A_2 = A_0(\Delta x/x_c)^{-\beta}$ with critical exponent $\beta$ = 0.72, showing that $A_2$ diverges via two orthogonal tuning parameters that both cooperate to direct the evolution of $B_c(x)$ through the $T$=0 field-doping plane (see Fig.1). The normalizing factor $A_0$ is equal to unity for 10 T data and scaled to unity for 0 T and 8 T value for the same reasons as above. **c** and **d**, Scaling plots of $\rho(T)$ of LCCO for $x$ = 0.15 and 0.17 in magnetic fields greater than $B_c$ showing that resistivity $\Delta\rho$ data divided by $A_2T^2$ collapse onto the same curve with a suitable choice of scaling exponent $\gamma$. The blue arrows indicate $\Delta B^\gamma/T_{FL}$, which delineates the Fermi liquid side with zero slope and ordinate equal to unity from the non Fermi liquid behavior with positive slope. The success of this scaling over two orders of magnitude in $\Delta B^\gamma/T$ indicates that the critical scaling of $A_2$ and the $\sim T^n$ resistivity have the same origin, and that magnetic field and temperature are the dominant energy scales in the system.

**Supplementary Information**

**Scaling of the quadratic coefficient $A_2$ of the resistivity in the Fermi liquid state.** Fits to the electrical resistivity of the form $\rho = \rho_0 + A_2T^2$ were performed on the high field side of $B_c$, i.e., in the field-induced Fermi liquid (FL) ground state. As a function of magnetic field, the data scale with identical critical exponent $\alpha = 0.38$. This analysis was performed on data from multiple samples at each concentration. In order to plot the data together in Fig. 3a, the absolute values of the coefficients for each sample were scaled by a constant value, which maintains the integrity of the scaling analysis. The necessity for rescaling is expected because of the sensitivity of the scattering to sample dependence beyond experimental control, which makes the success of the $A_2$ scaling all the more remarkable. A similar approach was used to put together Fig. 3b. Note that, experimentally, the step size is much coarser in the doping direction and the uncertainty is larger due to the aforementioned sample dependence.

**Resistivity scaling above the Fermi liquid boundary $B_c$.** The scaling of $\rho(T)$ reflects the fact that the resistivity $\Delta\rho$ can be described generally as a function $A_2T^2 f(\Delta B^\gamma/T)$, where $\Delta B = B - B_c$, that is applicable to scattering in both Fermi liquid ($\Delta\rho \propto T^2$) and non Fermi liquid ($\Delta\rho \propto T^n$) regions. In this framework, the $T^n$ behavior in the NFL region stems from anomalous temperature dependence in $A_2$, which is by definition a constant in temperature in the FL state. The resultant picture is that $T_{FL}$ separates the FL state at high magnetic field and low temperature from the NFL region at low magnetic field and high temperature, consistent with the magnetic field dependence of $T_{FL}$ (Fig. S1c). This also suggests that upon crossing $T_{FL}$ the dominant energy scale is transferred from temperature to magnetic field, which implicitly suggests that any dominant energy scale, such as a Fermi energy, is absent.

The exponents $\alpha$, $\gamma$, and $n$ are related as $\alpha = \gamma(2 - n)$ by the considering the following limits:

    1) Fermi liquid ($T << \Delta B$):          $\Delta\rho = A_2(B)T^2$

In this limit, $\Delta\rho / A_2(B)T^2 = 1$ and thus $f(\Delta B^\gamma/T) \rightarrow 1$.

    2) Non Fermi liquid ($T >> \Delta B$):    $\Delta\rho = A_n T^n = A_2(B)T^2 \times (\Delta B^\gamma/T)^{n-2}$

When $n < 1$, $\Delta\rho < A_2(B)T^2$ and $f(\Delta B^\gamma/T) < 1$. Note that it is possible to define $A_2'(B,T) = A_2(B) \times (\Delta B^\gamma/T)^{n-2}$, or in other words, explicitly add a temperature dependence to $A_2$, which is a constant in temperature in the Fermi liquid state. However, from Fig. 3a we already know that $A_2 \propto \Delta B^{-\alpha}$, and because of $A_2$ and $A_2'$ must have the same magnetic field dependence, it follows that $\gamma(n-2) = -\alpha$.

For $x = 0.17$, scaling is satisfied using an exponent $\gamma = 1.0\pm0.02$, so $\alpha = 0.38$ forces $n \approx 1.6$. For $x = 0.15$, scaling is satisfied using an exponent $\gamma = 0.4\pm0.1$, so $\alpha = 0.38$ forces $n \approx 1.0$ (see Figure S5).

The plots in Figure 3 show the difference between Fermi liquid and non Fermi liquid behavior. In the Fermi liquid state, $\Delta\rho/A_2T^2 = 1$ by definition and the slope of the scaled curve is zero. In contrast, in the non Fermi liquid regime the slope of the scaled curve is positive, reflecting the notion that $A_2$ is no longer a constant.

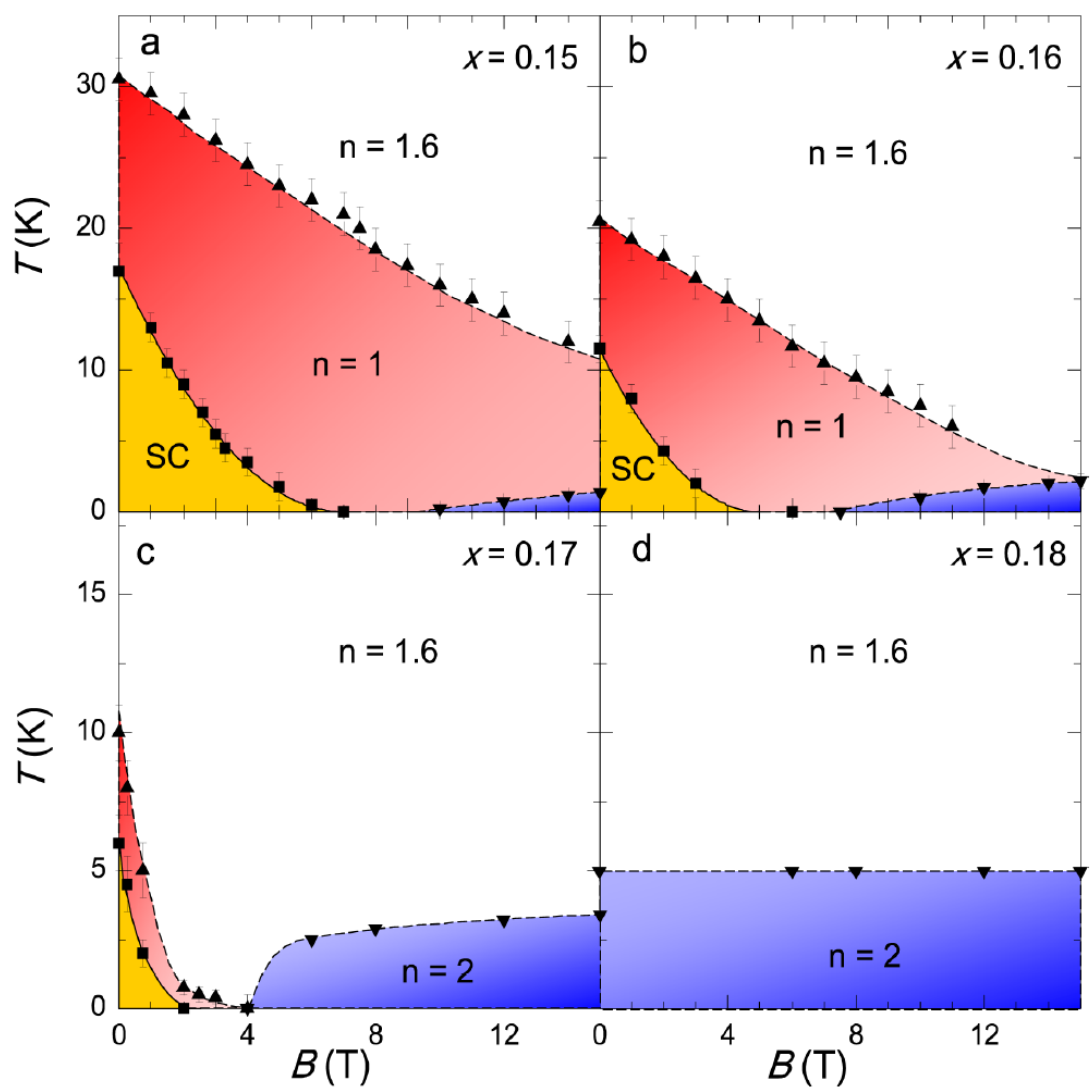

**Figure S1| LCCO magnetic field-temperature phase diagrams for *x* = 0.15, 0.16, 0.17 and 0.18.** These panels present constant-doping cuts of the phase diagram in Figure 1, illustrating the concave shape of the superconducting upper critical field curves (boundary of yellow regions) and the crossover between $\Delta\rho \propto T$ spin fluctuation-dominated scattering (red region) and $\Delta\rho \propto T^{1.6}$ scattering behavior. The evolution of the Fermi liquid state (blue regions) with $\Delta\rho \propto T^2$ is also shown to evolve with doping toward a dominant, field-independent state at *x* = 0.18.

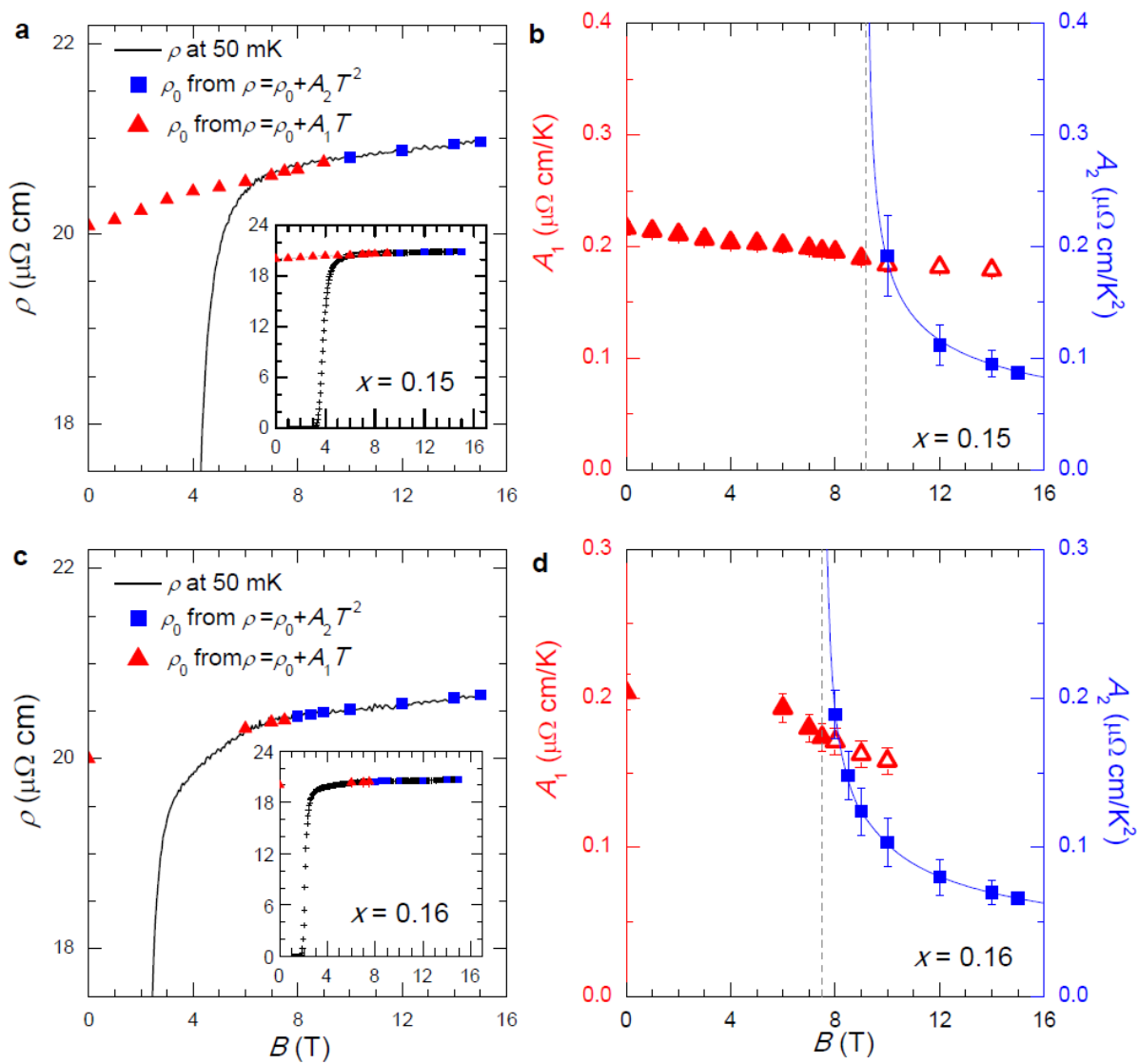

**Figure S2 | The magnetic field dependence of resistivity $\rho$ in the low temperature limit.** Panels **a** and **c** present vertical zooms of magnetoresistance data measured at a constant temperature of 50 mK for $x = 0.15$ and $x = 0.16$, respectively, and the residual (T=0) resistivity $\rho_0$ obtained from extrapolated fits of $\rho(T)$ at different constant fields. Insets present the full vertical axis scales for each data set. Panels **b** and **d** present the field dependence of inelastic scattering coefficients $A_1$ (red) and $A_2$ (blue) for $x = 0.15$ and $x = 0.16$, respectively, demonstrating the persistence of finite-temperature $\Delta\rho \propto T$ scattering beyond the critical field (dashed line), which is the critical boundary of $\Delta\rho \propto T$ and $\Delta\rho \propto T^2$ regions at zero temperature. Note that while $A_2$ exhibits an upturn at the critical field, $A_1$ is completely insensitive to $B_c$.

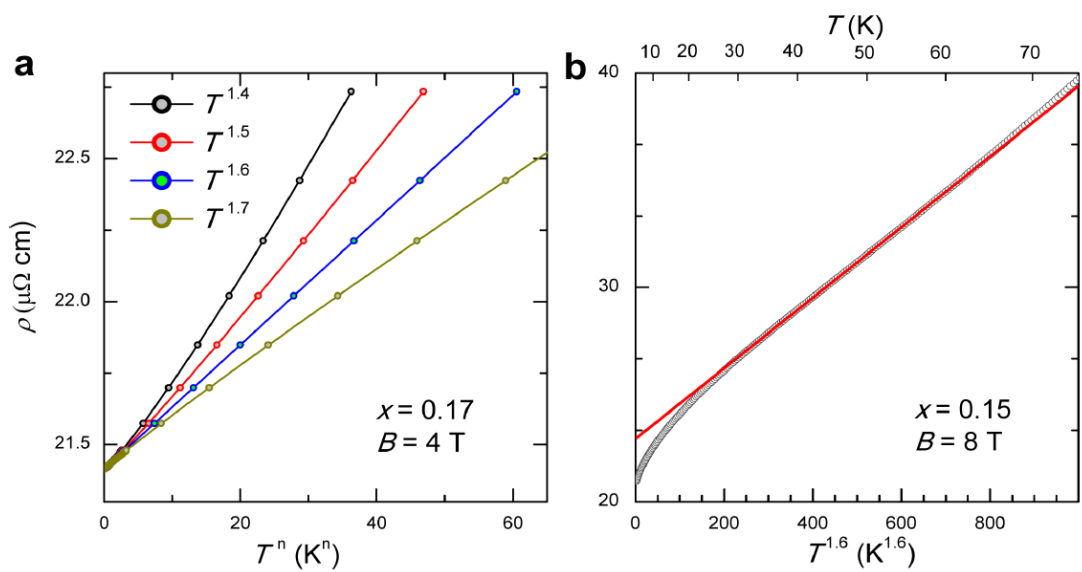

**Figure S3| Determination of $T^{1.6}$ resistivity. a**, Comparison of different power law temperature fits of the resistivity for $x$ = 0.17 in a magnetic field of 4 T, where both $\Delta\rho \propto T$ and $\Delta\rho \propto T^2$ behaviors vanish. It is clear that a $T^{1.6}$ power yields the best fit, as shown by the blue data. **b**, Demonstration of the range of the $\sim T^{1.6}$ power law fit for $x$ = 0.15 in a field of 8 T, where $\Delta\rho \propto T$ behavior is dominant from zero temperature up to a crossover temperature of ~ 20 K where the $\sim T^{1.6}$ power law becomes dominant and then extends up to ~60 K.

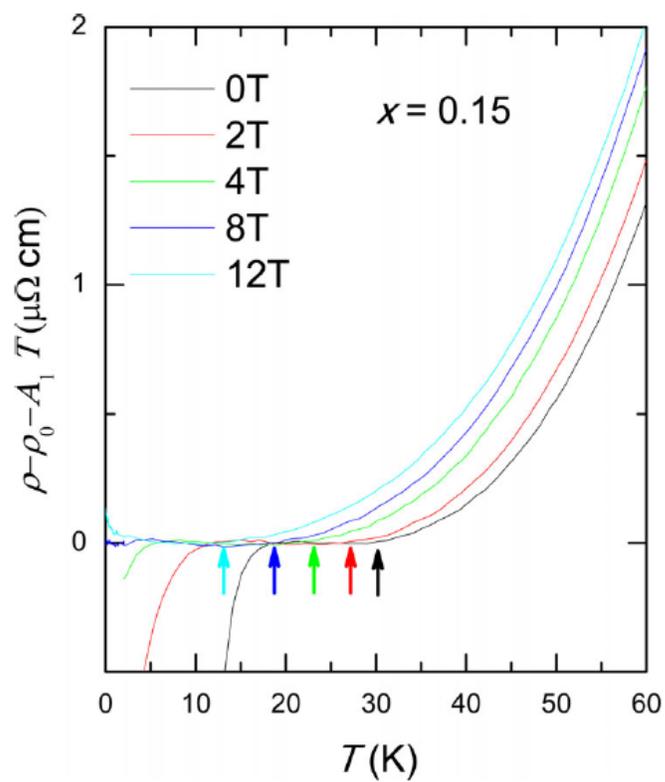

**Figure S4| Residuals of linear fits for *x* = 0.15.** The definition of $T_1$ is denoted by arrows. Of particular note are the 8 T data (blue line) where temperature-linear resistivity extends from 20 mK up to ~20 K.

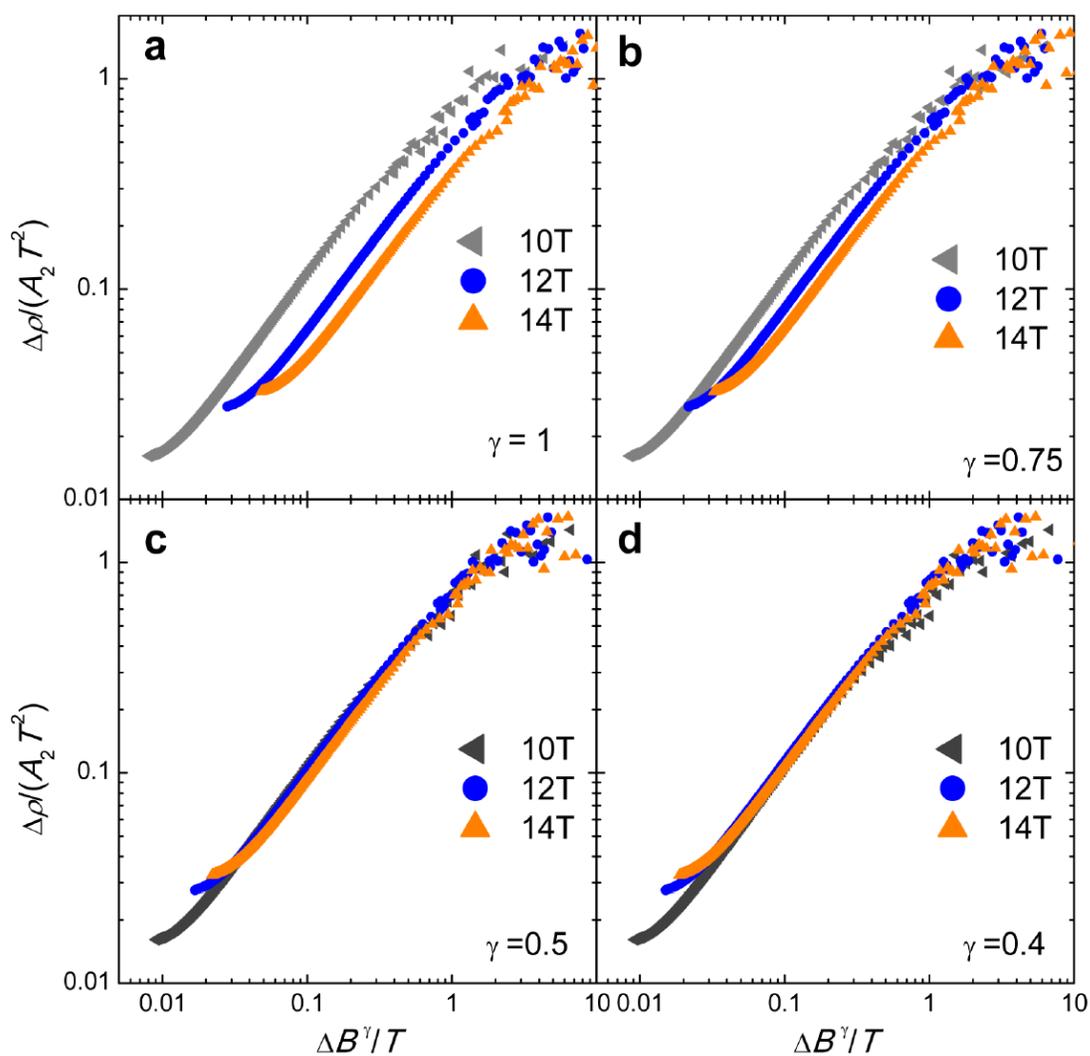

**Figure S5| Determination of resistivity scaling for $x = 0.15$.** This series of plots demonstrates that $\gamma$ should be considered as an independent fitting parameter and that the success of the scaling and its agreement with the critical scaling of $A_2$ as a function of $B$ is a demonstration of self-consistency between the exponents. For instance, it is clear in panel a) that the scaling exponent for x=0.15 is not 1.0 (*i.e.*, in contrast to the scaling observed for $x = 0.17$ with a choice of $\gamma=1.0$, as shown in Fig.3d).